**Scavenging of atmospheric ions and aerosols by drifting snow in Antarctica**

**A. K. Kamra[a], Devendraa Siingh[a,b]\* and Vimlesh Pant[a]**

[a] Indian Institute of Tropical Meteorology, Pune - 411 008, India

[b] Institute of Environmental Physics, University of Tartu, Ulikooli Street 18, 50090, Tartu, Estonia.

\*Corresponding Author:

    Dr. Devendraa Siingh
    Indian Institute of Tropical Meteorology,
    Dr Homi Bhabha Road,
    NCL Post, Pashan, Pune-411 008
    India

    Email: devendraasiingh@tropmet.res.in
       devendraasiingh@gmail.com
       devendraasiingh@yahoo.com

    Telephone: +91 -20-25893600 Ext 287

    Fax: +91-20-25893825

**Abstract**

   Measurements of the small-, intermediate-, and large-ion concentrations and the air-earth current density along with simultaneous measurements of the concentration and size-distribution of aerosol particles in the size-ranges 4.4 - 163 nm and 0.5 - 20 µm diameters are reported for a drifting-snow period after the occurrence of a blizzard at a coastal station, Maitri, Antarctica. Ion concentrations of all categories and the air-earth current simultaneously decrease by approximately an order of magnitude as the wind speed increases from 5 to 10 ms$^{-1}$. The rate of decrease is the highest for large ions, lowest for small ions and in between the two for intermediate ions. Total aerosol number concentration decreases in the 4.4 – 163 nm size-range but increases in 0.5 – 20 µm size range with wind speed. Size distribution of the nanometer particles show a dominant maximum at ~ 30 nm diameter throughout the period of observations and the height of the maximum decreases with wind speed. However, larger particles show a maximum at ~ 0.7 µm diameter but the height of the maximum increases with increasing wind speed. The results are explained in terms of scavenging of atmospheric ions and aerosols by the drifting snow particles.

**1. Introduction**

   Raising of snow/dust from the Earth's surface into the atmosphere by strong winds generates large charges and produces strong electrification in snowstorms and dust storms (e.g. Simpson, 1919; Latham, 1964; Kamra, 1969a, 1972a). Dispersion of snow/dust not only introduces ions into the atmosphere but also strongly modifies the fair-weather electrical state of the atmosphere (Kamra, 1969b). The electrification produced is sometimes strong enough to cause lightning/electric sparks in the atmosphere (Kamra, 1972b). However, not much is known about the processes causing the electrification or even about the type of ions being introduced into the atmosphere, Once dispersed into the atmosphere, these ions produce electrostatic effects, influence the conduction and convection currents flowing in the atmosphere and modify the electric state of the atmosphere in a variety of ways.

   The electrical state of the atmosphere is largely determined by the presence and movement of ions in the atmosphere. While the ions in the lower atmosphere are generated mainly by cosmic rays and radioactivity in the ground, they can also be generated by a variety of other mechanisms such as the bursting of air bubbles on the air-water interface of water bodies, splashing of raindrops on the Earth's surface, combustion activities, automobile exhausts, point discharge from the points raised above the Earth's surface etc. The ions generated and introduced in the atmosphere are destroyed by several processes such as recombination of ions, attachment to aerosol particles, sedimentation and scavenging by raindrops or ice particles. However, the contribution of such individual processes to the ionic equilibrium state of the atmosphere is not easily distinguishable one from the other. Measurements made over land stations are often largely disturbed by anthropogenic pollutants. The effects of individual factors are more effectively illustrated if such measurements are made at a clean place such as the open ocean, Antarctica, Arctic etc with almost no disturbance due to anthropogenic sources. An opportunity for such a study was provided during our measurements of some atmospheric electric parameters and aerosols during the Indian Scientific Expedition to Antarctica. Here, we report our measurements of different categories of atmospheric ions, air-earth current density and concentration and size distribution of aerosol particles after the occurrence of a blizzard at a coastal station, 'Maitri' in Antarctica.

## 2. Instrumentation

Concentrations of the small-, intermediate-, and large-ions are measured with an ion counter consisting of three different coaxial Gerdien condensers through which air is sucked with a single fan [Dhanorkar and Kamra, 1991; Siingh et al., 2005, 2007]. It can simultaneously measure small ($> 0.77 \times 10^{-4}$ $m^2V^{-1}s^{-1}$; $< 1.45$ nm size), intermediate ($1.21 \times 10^{-6} - 0.77 \times 10^{-4}$ $m^2V^{-1}s^{-1}$; $1.45 - 12.68$ nm size) and large ($0.97 \times 10^{-8} - 1.21 \times 10^{-6}$ $m^2V^{-1}s^{-1}$; 12.68 to ~ 130 nm size) ions. The air-earth current density is measured with a 1 $m^2$ flat-plate antenna installed flush with the ground. Concentration and size distribution of aerosol particles were measured with a Scanning Mobility Particle Sizer (SMPS, Model 3936 of TSI.) system in the size-range 4.4 - 163 nm diameter and with an Aerodynamic Particle Sizer (APS, Model 3321 of TSI.) system in the size-range $0.5 - 20.0$ µm diameter. Sensors for the atmospheric electricity measurements were installed ~ 15 m away from a hut and the samples for the aerosol measurements were sucked through a tube fixed in a wall of the hut. All measurements were performed continuously and recorded and stored with a data logger and computer kept inside the air-conditioned hut. Details of the instruments and their installation are given in Siingh et al. (2007).

All instruments were installed about 300 m away and to the East of the station so that the measurements are not likely to be polluted by the exhausts from the station in prevailing southeasterly winds at the station.

## 3. Weather

A blizzard on February 18-19, 2005 accompanied with snowfall deposited snow all over the 'Maitri' Antarctica. All measurements were discontinued during this period but were started on the morning of February 20, 2005. when the weather was clear. Sunlight was intense after 1000 UT. Figure 1 shows the variations of meteorological parameters at Maitri on February 20. Southeasterly winds, though high throughout the day, sometimes exceeded 10 $ms^{-1}$ after 1700 UT and caused some blowing snow in the lower atmosphere. Atmospheric temperatures varied between -1 to -5°C. There was not much change in atmospheric pressure which remained almost constant at 960 - 961 hPa. However, cloud coverage increased and became almost overcast after 2000 UT.

## 4. Observations

### (a) Ion and air-Earth current measurements

Figure 2 shows the variations in concentrations of ions of three categories and the air-Earth current density from 1200 to 2400 UT on February 20, 2005. Winds are not very strong and generally remain < 6 ms$^{-1}$ up to 1700 UT. Concentrations of the small-, intermediate,- and large-ions also remain in the ranges 2 to 4 × 10$^8$, 8 × 10$^9$ to 2 × 10$^{10}$ and 7 × 10$^9$ to 1.5 × 10$^{10}$ m$^{-3}$ respectively up to 1700 UT and do not show large variations during this period. The corresponding air-Earth current density during this period varies from 0.4 to 0.8 × 10$^{-12}$ A m$^{-2}$. However, ion concentrations of all the three categories start decreasing after 1800 UT and the decrease is by more than an order of magnitude in the next 1 to 3 hours as the wind speed increases from 5 to 10 ms$^{-1}$. The rate of decrease is the highest for the large ions, the lowest for small ions and in between the two for intermediate ions. The concentrations of small-, intermediate- and large-ions decrease from their ambient values to the minimum values of 2 × 10$^7$, 10$^8$ and 3 × 10$^8$ m$^{-3}$ in 2 h 53 min, 1 h 54 min, and 1 h 14 min, respectively. The air-earth current density also decreases from 0.7 pAm$^{-2}$ to almost zero in 1 h 26 min.

**(b) Aerosol measurements**

**(i) Total aerosol number concentrations**

Figure 3 shows that the total number concentration of aerosol particles in the size-range of 4.4 – 163 nm diameters decreases from ~ 1800 to ~ 800 cm$^{-3}$ as the wind speed increases from 5 to 10 ms$^{-1}$ between 1700 and 2200 UT. However, on the contrary, the total number concentration of particles in the size-range 0.5 to 20 µm diameter increases from 0.2 to 0.4 cm$^{-3}$ in the same time period.

**(ii) Number size distributions**

Figure 4 shows size distributions of nanometer particles and their evolution at an interval of every 2 hours. All curves show a dominant maximum at ~ 30 nm diameter and the height of maximum decreases as the wind speed increases from 1700 to 2200 UT. However, another maximum appears at ~ 11 nm diameter in the afternoon hours when the solar radiation is high. Moreover, the particle size distributions obtained at these hours show the presence of particles as small as 4.6 nm diameter. Since such small particles can not be transported from long distances, their presence indicates the formation of new particles of nanometer size near this locality and their subsequent growth to these sizes (Koponen et al. 2003; Kulmala and Kerminen, 2008).

Figure 5 shows the evolution of the size distributions of the particles in the size range 0.5 – 20 µm diameter. All size distributions show a broad maximum between 0.7 and 1.0 µm diameter. However, contrary to the trend of the maxima height decreasing with the wind speed in case of the nucleation and accumulation modes, the height of maxima in this coarse particle mode increases with the increase in wind speed.

**5. Discussion**

As a result of a blizzard on February 18-19, 2005, the whole area around Maitri was almost covered with snow. Strong winds blowing over this loose snow on February 20 were observed to raise snow particles which drift with the horizontal winds. Our observations of different rates of decrease in the case of different categories of ions in Figure 2 can be explained in terms of the scavenging of atmospheric ions and aerosols by the drifting snow particles. The small ions generated close to the ground are soon attached to the aerosol particles of different sizes. As a result of the difference in size of the blowing snow particles and atmospheric ions of different categories, snow particles and ions will develop a relative velocity. While ions of all size-categories will be carried by the air, snow particles will experience a drag force and attain somewhat lower velocities. This difference in velocities of snow particles and ions will cause the scavenging of ions by snow particles. Now, collection efficiencies of particles of the size of blowing snow are 3 – 4 orders of magnitude higher for particles of the size of large ions than for small ions (Wang et al., 1978). So the blowing snow particles will collect the large ions much more efficiently than smaller ions. In addition, large ions, because of having larger terminal velocities, will quickly settle to the ground by sedimentation. This may be the cause of the higher rate of decrease of large ions compared to that of small ions in our observations.

The concentration of intermediate ions may be influenced by yet another process. Covert et al. (1996) and Ito (1985) have shown that the formation of new particles by the photo-oxidation process can occur in a clean environment with low aerosol surface area if plenty of solar radiation is available. The rate of new particle formation is strengthened by the higher emissions of dimethyl sulphide (DMS) in the ice-melt regions around the continent of Antarctica (Davison et al., 1996; O'Dowd et al., 1997). Enhanced solar radiation in the afternoon hours at 1600 and 1700 UT may causes higher formation rates of new particles by the gas-to-particle conversion processes and can explain comparatively higher maxima at both hours in the size distributions of nano-size particles. The small ions generated by cosmic rays can get attached to these particles or enhance the new particle formation process by ion nucleation.

**Acknowledgements**

The authors express their gratitude to the National Centre for Antarctic and Ocean Research (NCAOR) and the Department of Ocean Development (DOD) for participation in the 24$^{th}$ Indian Antarctic Expedition. The meteorological data provided by India Meteorological Department is thankfully acknowledged. DS thanks Department of Science and Technology (DST), Government of India for award the BOYSCAST fellowship and Dr Urmas Horrak and Mrs. Karin Tuvikene, Institute of Environmental Physics, University of Tartu, Estonia for their help during the visit.

The authors thank the two anonymous reviewers for constructive and valuable suggestions.

**Figure Captions**

1. Variations of meteorological parameters measured on February 20, 2005 at Indian station Maitri, Antarctica.

2. Concentrations of small-, intermediate-, and large-positive ions and air-earth current density on February 20, 2005.

3. Total aerosol number concentration in the size range 4.4 – 163 nm and 0.5 – 20 µm diameter along with wind speed on February 20, 2005.

4. Aerosol number size distributions in the size range 4.4 – 163 nm at an interval of 1 hour on February 20, 2005.

5. Aerosol number size distributions in the size range 0.5 – 20 µm at an interval of 1 hour on February 20, 2005.

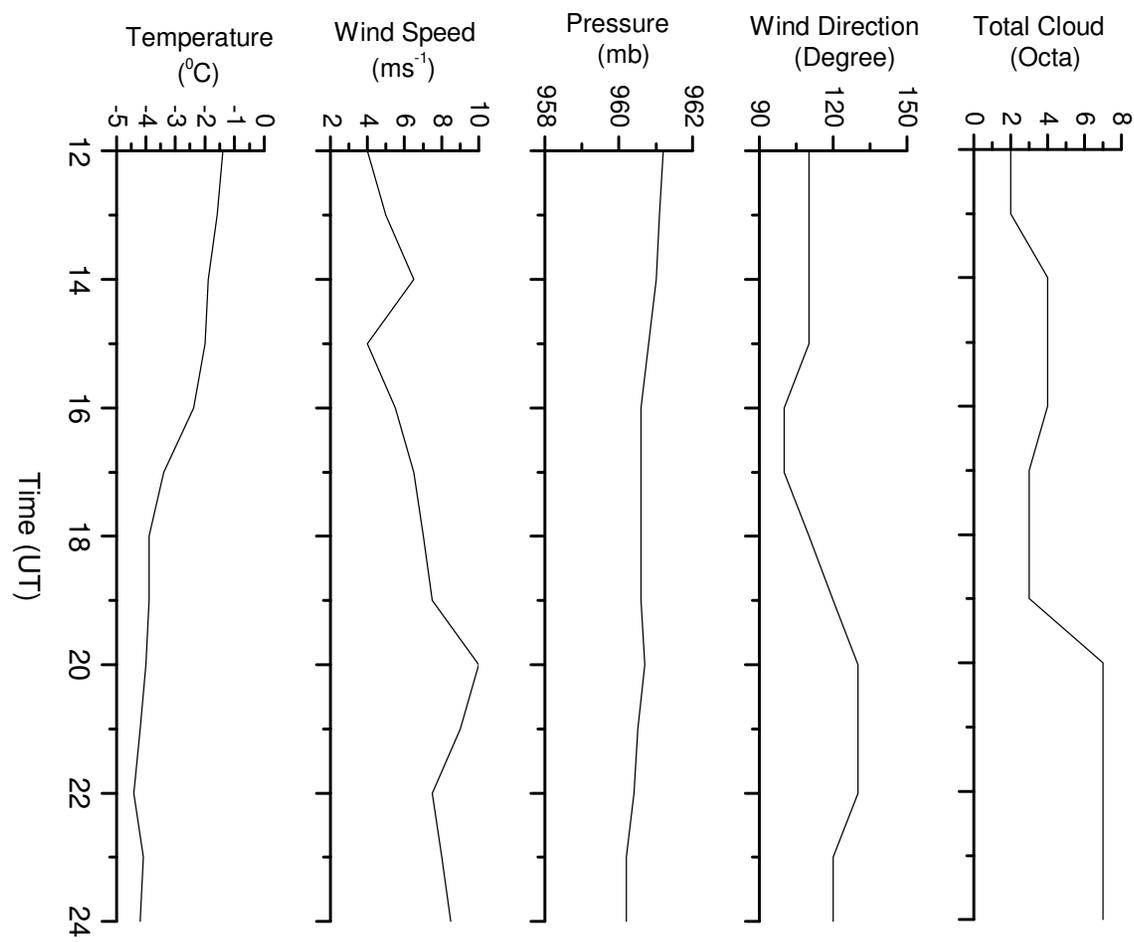

Fig1

Fig 2

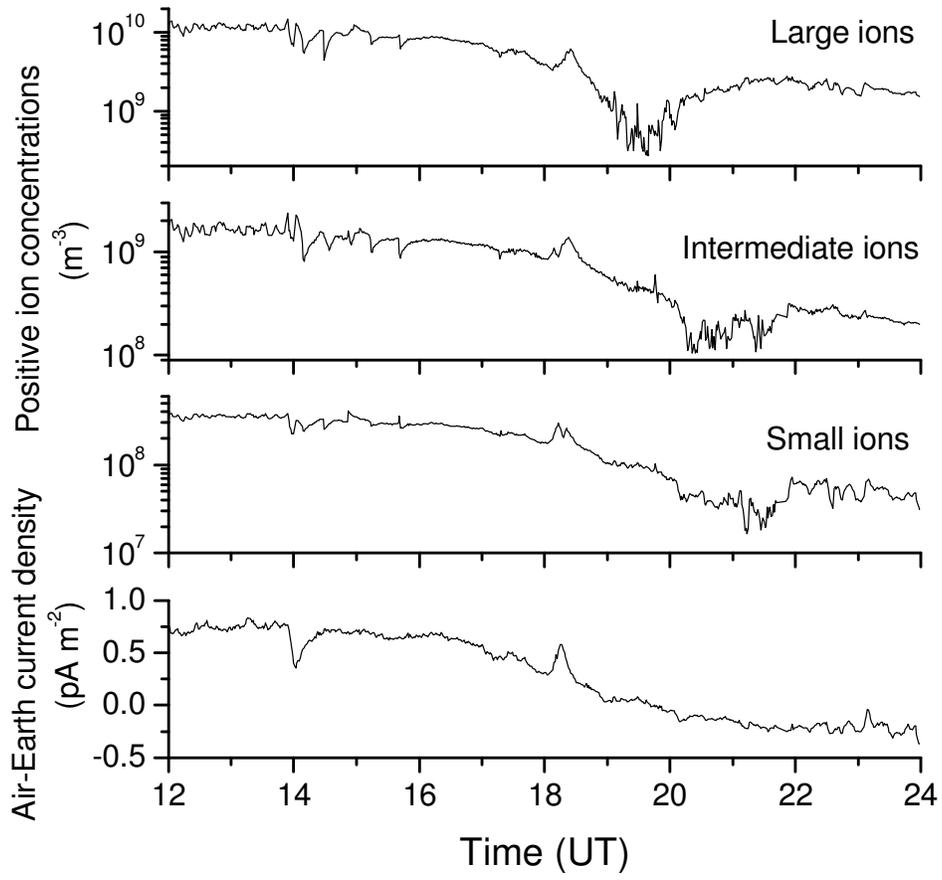

Fig 3

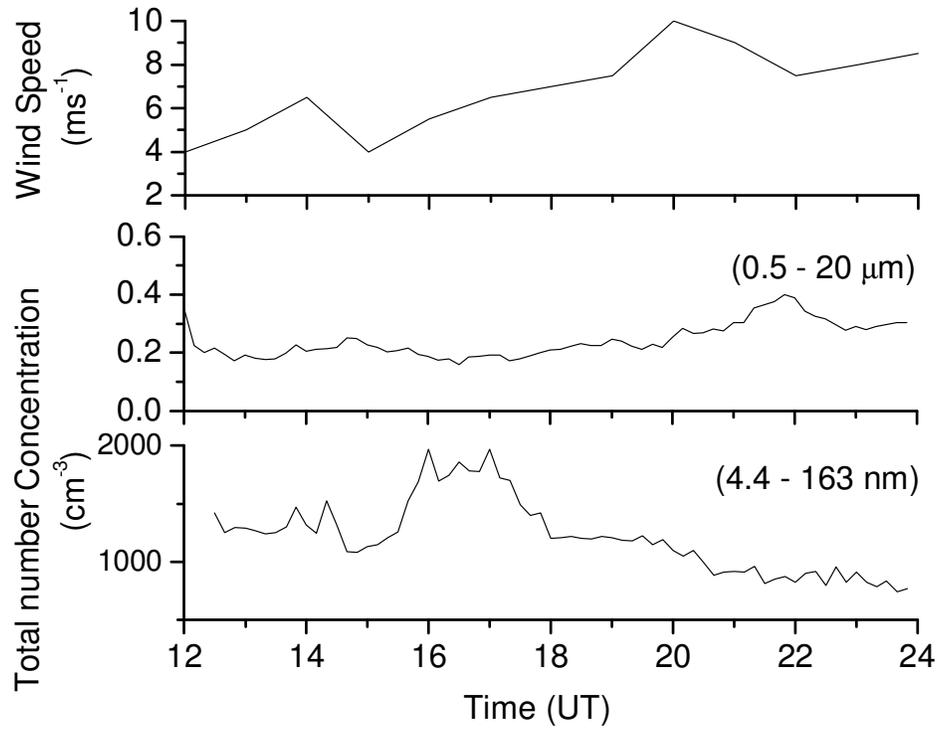

Fig 4

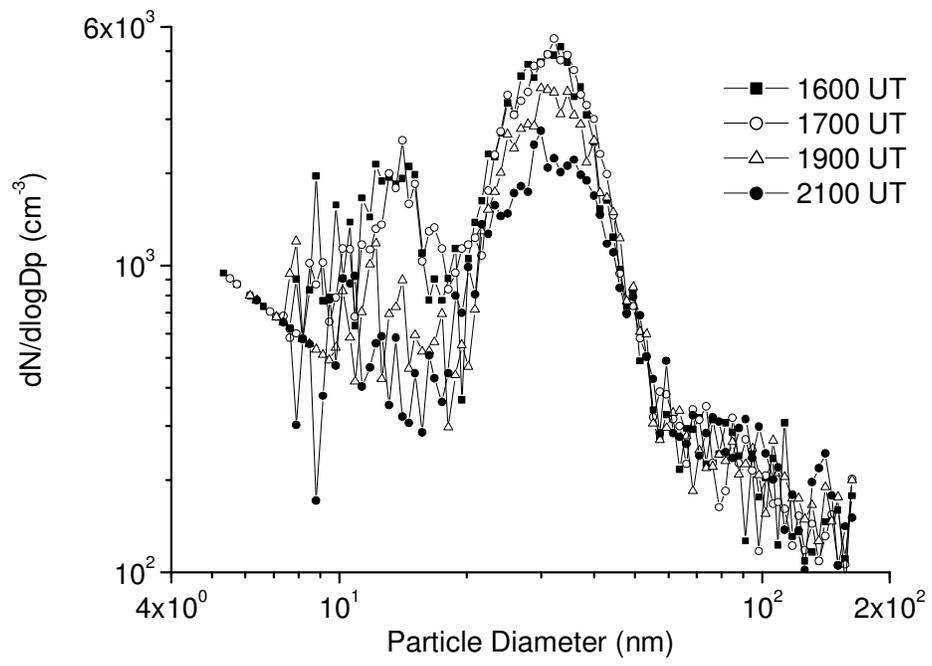

Fig 5

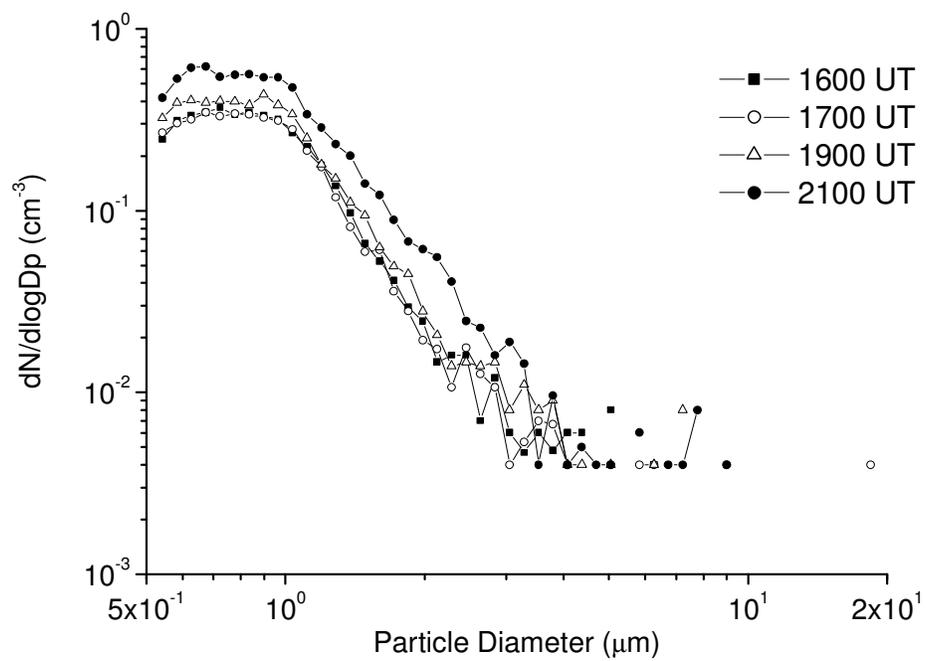